\documentclass[12pt,amssymb,amsmath,english,aps,preprint,showpacs]{revtex4}
\usepackage{hyperref}
\usepackage{subfigure}
\usepackage{graphicx}

\renewcommand{\b}[1]{\boldsymbol{#1}}

\newcommand{\unit}[1]{\,{\rm #1}}
\newcommand{\cm}{\unit{cm}}
\newcommand{\G}{\unit{G}}
\newcommand{\kG}{\unit{kG}}
\newcommand{\rpm}{\unit{rpm}}
\newcommand{\s}{\unit{s}}
\newcommand{\ab}{{\rm a}}
\newcommand{\bp}{{\rm b}}

\newcommand{\half}{{\textstyle\frac{1}{2}}}

\newcommand{\Rm}{Re_{\rm m}}

\newcommand{\we}{\omega_\eta}
\newcommand{\wo}{\omega_{\textsc{IO}}}
\newcommand{\wt}{\omega_\theta}
\newcommand{\wz}{\omega_z}

\newcommand{\pd}{\partial}

\providecommand{\boldsymbol}[1]{\mbox{\boldmath $#1$}}

\providecommand{\tabularnewline}{\\}

\usepackage{babel}

\begin{document}

\preprint{APS/123-DPP}

\title{Helical Magnetorotational Instability in Magnetized Taylor-Couette Flow}

\author{Wei Liu$^{1}$\footnote{Email: wliu@pppl.gov}, Jeremy Goodman$^{2}$, Isom Herron$^{3}$, Hantao Ji$^{1}$}

\affiliation{$^{1}$Center for Magnetic Self-Organization in Laboratory and Astrophysical Plasma, Princeton Plasma Physics Laboratory, Princeton University, P.O. Box
451, Princeton, NJ 08543 }

\affiliation{$^{2}$Princeton University Observatory, Princeton, NJ 08544}

\affiliation{$^{3}$Department of Mathematical Sciences, Rensselaer
Polytechnic Institute, Troy, NY 12180}

\date{\today}

\begin{abstract}
  Hollerbach and R\"udiger have reported a new type of
  magnetorotational instability (MRI) in magnetized Taylor-Couette
  flow in the presence of combined axial and azimuthal magnetic
  fields.  The salient advantage of this ``helical'' MRI (HMRI) is
  that marginal instability occurs at arbitrarily low magnetic
  Reynolds and Lundquist numbers, suggesting that HMRI might be easier
  to realize than standard MRI (axial field only). We confirm their
  results, calculate HMRI growth rates, and show that in the resistive
  limit, HMRI is a weakly destabilized inertial oscillation
  propagating in a unique direction along the axis. But we report
  other features of HMRI that make it less attractive for experiments
  and for resistive astrophysical disks.  Growth rates are small and
  require large axial currents.  More fundamentally, instability of
  highly resistive flow is peculiar to infinitely long or periodic
  cylinders: finite cylinders with insulating endcaps are shown to be
  stable in this limit.  Also, keplerian rotation profiles are stable
  in the resistive limit regardless of axial boundary conditions. Nevertheless, the addition of toroidal field
  lowers thresholds for instability even in finite cylinders.

\end{abstract}


\maketitle

\section{Introduction}\label{sec:intro}
The magnetorotational instability (MRI) is probably the main source of
turbulence and accretion in sufficiently ionized astrophysical disks
\cite{bh98}. MRI was first discovered theoretically
\cite{ve59,chan60,bh91}, then later supported numerically
\cite{hgb95,bnst95,mt95}, but has never been directly observed in
astronomy.  No laboratory study of MRI has been completed except for
that of \citet{sisan04}, whose experiment proceeded from a background
state that was not in MHD equilibrium.  We and others therefore have
proposed experimental demonstrations of MRI \cite{jgk01,gj02,npc02}.
The experimental geometry planned by most groups is a magnetized
Taylor-Couette flow: an incompressible liquid metal confined between
concentric rotating cylinders, with an imposed background magnetic
field sustained by currents external to the fluid.

The challenge for experimentation, however, is that liquid-metal flows
are very far from ideal on laboratory scales.  While the fluid
Reynolds number $Re\equiv \Omega_{1}r_{1}(r_{2}-r_{1})/\nu$ can be
large, the corresponding \emph{magnetic} Reynolds number
$\Rm\equiv\Omega_{1}r_{1}(r_{2}-r_{1})/\eta$ is modest or small,
because the magnetic Prandtl number $Pr_{\rm m}\equiv\nu/\eta\sim
10^{-5}-10^{-6}$ in liquid metals; here $\nu\lesssim
10^{-2}\cm^2\s^{-1}$ is the kinematic viscosity and $\eta$ is the
magnetic diffusivity.  Standard MRI modes will not grow unless both
the rotation period and the Alfv\'en crossing time are shorter than
the timescale for magnetic diffusion.  This requires both $\Rm\gtrsim
1$ and $S\gtrsim 1$, where $S\equiv V_{A}(r_{2}-r_{1})/\eta$ is the
Lundquist number, and $V_{A}=B/\sqrt{\mu_0\rho}$ is the Alfv\'en speed.
Therefore, $Re\gtrsim 10^6$ and fields of several kilogauss must 
typically be achieved.

Recently, Hollerbach and collaborators have discovered that MRI-like
modes may grow at much reduced $\Rm$ and $S$ in the presence of a
helical background field, a current-free combination of axial and
toroidal field \citep{hr05,rhss05}.
\begin{equation}
  \label{eq:backgroundfield}
  \b{B}^{(0)}=B_z^{(0)}\left(\b{e}_z+\beta\frac{r_1}{r}\b{e}_\theta\right)
\end{equation}
in cylindrical coordinates $(r,\theta,z)$, where $B_z^{(0)}$ and
$\beta$ are constants.  (When it will not cause ambiguity, we will
omit the superscript $(0)$ from $\b{B}$ and $B_z$ hereafter.)
Henceforth, ``standard MRI'' (SMRI) will refer to cases where the
$\beta=0$, and ``helical MRI'' (HMRI) to modes that require
$\beta\ne0$.  In centrifugally stable flows---meaning that
$d(r^2\Omega)^2/dr>0$, where $\Omega=V_\theta^{(0)}/r$ is the
background angular velocity---SMRI exists only when $\Rm$ and $S$
exceed thresholds of order unity \cite{jgk01,gj02}.  Remarkably,
however, HMRI may persist in such flows even as both parameters tend
to zero, though not independently: more precisely, the thresholds are
$\ll 1$ and would vanish if the fluid were inviscid ($\nu=0$).  In a
fixed geometry and flow profile, the resistive limit may be approached 
theoretically by increasing $\eta$ with all other parameters held constant. The
growth rate of inviscid HMRI is then $\propto\eta^{-1}$ so that the
hydrodynamic case is approached continuously.
The special case of
toroidal-only magnetic field ($\beta =\infty$)
is stable \cite{hs06}.

%
%
%

A novel feature of the background state for HMRI is that there is a
uniform axial flux of angular momentum carried by the field,
$rT_{\varphi z}^{(\rm mag)}=-rB_\theta B_z/\mu_0$ and an associated
axial Poynting flux $\Omega$ times this.  In an infinite or periodic
cylinder, the question of the sources and sinks of these axial fluxes
need not arise, but in an experimental device, a torque is exerted by
the axial field on the radial sections of the coil that complete the
circuit containing the axial current.  Related to this perhaps, the
dispersion relation for linear modes is sensitive to the sign of the
axial wavenumber ($k_z$), and the instabilities of axially infinite or
periodic cylinders are travelling rather than standing waves, as noted
by Knobloch \citep{ke92,ke96}.  This begs the question what should
happen to the modes in finite cylinders, a question that has motivated
much of our analysis.

Even the analysis for periodic cylinders implies two practical difficulties
for an HMRI experiment.  First, as will be seen, the typical growth
rates tend to be smaller than those of SMRI except in regimes where SMRI
would also be unstable.  Secondly, the axial current needed for
the required toroidal fields tend
to be quite large: $I[\unit{kA}]=5 B_\theta r[\mbox{kG-cm}]$.

In Section \ref{sec:theory} we analyze the linear stability of HMRI
using complementary approximations, some for infinite/periodic
cylinders and others for finite ones.  The results are compared with
one another and with fully nonlinear axisymmetric simulations.
Our conclusions are summarized in Section \ref{sec:dis}.

\section{Linear theory}\label{sec:theory}

All magnetic fields are expressed as Alfv\'en
speeds, in other words, units such that $\mu_0=1/\rho$ are used.
Upper-case letters are used for the background magnetic field
(\ref{eq:backgroundfield}) and velocity
$\b{V}=r\Omega(r)\b{e}_\theta$, and lower-case ($\b{b},\b{v}$) for
perturbations.  Frequently occurring derivatives are abbreviated by
$\pd_r^\dag\equiv\pd_r+r^{-1}$, $D\equiv\pd_r\pd_r^\dag+\pd_z^2$.
Incompressibility allows the use of
stream functions for the poloidal components: $v_r=\pd_z\phi$,
$v_z=-\pd_r^\dag\phi$, $b_r=\pd_z\psi$, $b_z=-\pd_r^\dag\psi$; note
that these definitions differ by factors of $r$ from the usual ones.
The linearized inviscid MHD equations then become, since
$B_z$ and $rB_\theta$ are constant,
\begin{align}
\label{polind}
(\underline{\pd_t}-\eta D)\psi &= B_z\pd_z\phi,\\
\label{torind}
(\underline{\pd_t}-\eta D)b_\theta &= \pd_z\left(\frac{2B_\theta}{r}\phi+B_z v_\theta
+\underline{r\Omega'\psi}\right),\\
\label{poleuler}
\pd_t D\phi -2\Omega\pd_z v_\theta &= B_z\pd_z D\psi -\frac{2B_\theta}{r}\pd_z b_\theta,
\\
\label{toreuler}
\pd_t v_\theta+r^{-1}(r^2\Omega)'\pd_z\phi &= B_z\pd_z b_\theta.
\end{align}
The underlined terms above are
negligible in the resistive limit, where $\b{b}$
scales $\propto\eta^{-1}$ compared to $\b{v}$.
Neglecting these terms has been shown to suppress SMRI \cite{gj02,hg06}, 
but not HMRI as will be seen.

Taking another time derivative of \eqref{poleuler} and eliminating
$\pd_t v_\theta$ \emph{via} \eqref{toreuler} yields
\begin{equation}
  \label{phi1}
\left(\pd_t^2 D+\kappa^2\pd_z^2\right)\phi=
 B_z\pd_z\pd_t D\psi+2\left(\Omega B_z\pd_z^2
-\frac{B_\theta}{r}\pd_z\pd_t\right)b_\theta\,,
\end{equation}
in which $\kappa^2\equiv r^{-3}d(r^2\Omega)^2/dr^2$ is the square
of the epicyclic frequency.
As $\eta\to\infty$, \eqref{toreuler} reduces to
\begin{equation}\label{ino0}
\left(\pd_r\pd_r^\dag+\pd_z^2\right)\pd_t^2\phi~+\kappa^2(r)\pd_z^2\phi=0.
\end{equation}

\subsection{WKB for infinite or periodic cylinders}\label{sec:wkb}

If we take the gap to be narrow, $d\equiv r_2-r_1\ll r$, then it is
reasonable to treat $r,B_z,\Omega$, $r\Omega'=2Ro\Omega$, and
$r^{-1}(r^2\Omega)'=2(1+Ro)\Omega=\kappa^2/2\Omega$ as constants, and
to look for perturbations $\propto\exp(ik_r r+ik_z z-i\omega t)$.
The Rossby number $Ro\equiv \half d\ln\Omega/d\ln r$ has been introduced.
In this case one
expects to have WKB solutions with $D$ replaced by $-(k_r^2+k_z^2)\equiv
K^2$, where the total wavenumber $K=O(d^{-1})$.

When applied to \eqref{ino0} (\emph{i.e.} for $\eta\to\infty$) these
prescriptions yield the dispersion relation for hydrodynamic inertial
oscillations (hereafter IO),
\begin{equation}\label{IO}
\wo^2= \kappa^2\frac{k_z^2}{k_r^2+k_z^2}\quad\mbox{where}\quad
\kappa^2=\frac{1}{r^3}\frac{d}{dr}(r^2\Omega)^2 = 4(1+Ro)\Omega^2.
\end{equation}
IO exist only in the Rayleigh-stable regime $\kappa^2>0$,
$Ro>-1$, and their frequencies lie between $0$ and $\kappa$.

HMRI occurs at finite $\eta$ when $B_\theta/r \equiv\wt$ is comparable
to $k_{z}B_z\equiv\wz$.  Define $\we\equiv\eta K^2$ and $\mu\equiv
k_{z}/|K|\in[-1,1]$.  The dispersion relation corresponding to the
system \eqref{polind}-\eqref{toreuler} is then
\begin{multline}\label{fulldisp}
0=s^4+2\we s^3+\left[\we^2+4\mu^2\wt^2+2\wz^2+\mu^2\kappa^2\right]s^2\\
+2\left[2\we\mu^2\wt^2+\we\wz^2+\we\mu^2\kappa^2-4i\mu^2\wt\wz\Omega\right]
s\\
+\left[\we^2\mu^2\kappa^2-4i\we\wt\wz\mu^2\Omega(2+Ro)+\wz^4
+4\mu^2\wz^2\Omega^2 Ro\right],
\end{multline}
where the complex growth rate $s\equiv -i\omega$ 
has been used so that the coefficients are
all real except for those linear in $\wt$.
It is instructive to consider the limit in which
$\omega_\eta$ is much larger than all of the other frequencies,
including $\omega$:
\begin{equation}\label{disp2}
s^2+\wo^2~+2\we^{-1}\left[s^3
+(2\mu^2\wt^2+\wz^2+\wo^2)s
-2i\wt\wz\mu^2\Omega(2+Ro)\right]
~\approx~O(\we^{-2}).
\end{equation}
The replacement $\mu^2\kappa^2\to\wo^2$ emphasizes that
$\omega\approx\pm\wo$ in this limit.  The roots are
\begin{equation}\label{roots}
\omega\approx \mp \wo~+i\we^{-1}\left[\pm2\wt\wz\wo^{-1}\mu^2\Omega(2+Ro)~-
\left(2\mu^2\wt^2+\wz^2\right)\right]~+O(\we^{-2}),
\end{equation}
the bivalent signs being correlated.
The other two roots of \eqref{fulldisp} represent rapidly decaying magnetic perturbations,
$s\approx -\we$.

\begin{figure}[!htp]

\caption{Selected roots of full dispersion relation \eqref{fulldisp} for 
$\eta=2,000\cm^2\s^{-1}$ [gallium], 
$r_1=9\cm$, $r_2=11\cm$, vertical periodicity
$2h=16\cm$, $\Omega_1=100\rpm$, 
$\Omega_2=68.1\rpm$, $B_{z}=500\G$, $B_{\theta}=10\kG$ at $r=(r_1+r_2)/2$.
The two rapidly damped modes are omitted.\label{dispersion} }

\subfigure[Growth rate $\gamma=\Im{\omega}$ vs. Wave number
$k_{z}$]{\scalebox{.4}{\includegraphics[
    ]{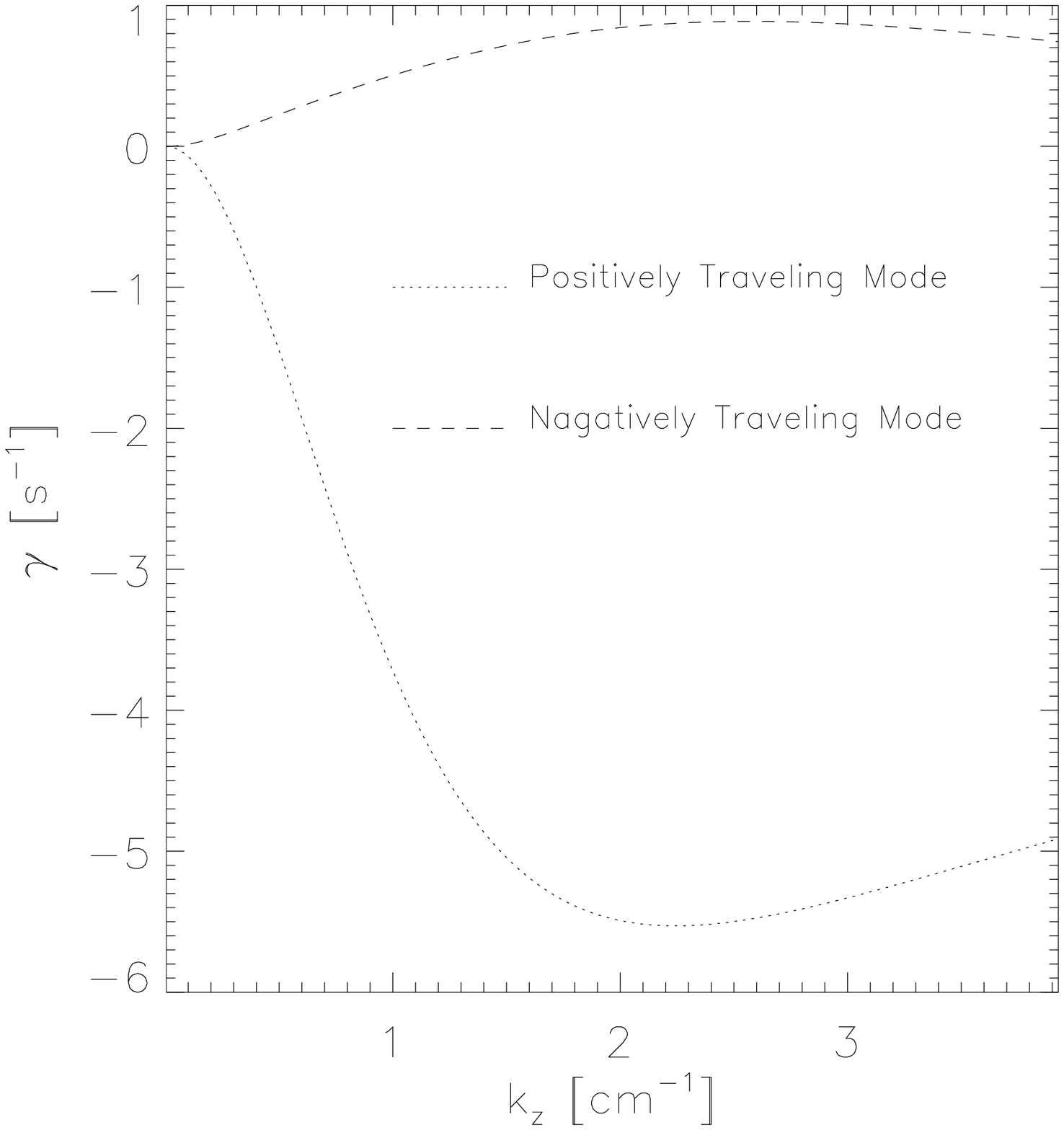}}}$\;$\subfigure[Real frequency $\omega_{r}=\Re{\omega}$ vs.
Wave Number $k_{z}$]{\scalebox{.4}{\includegraphics[
    ]{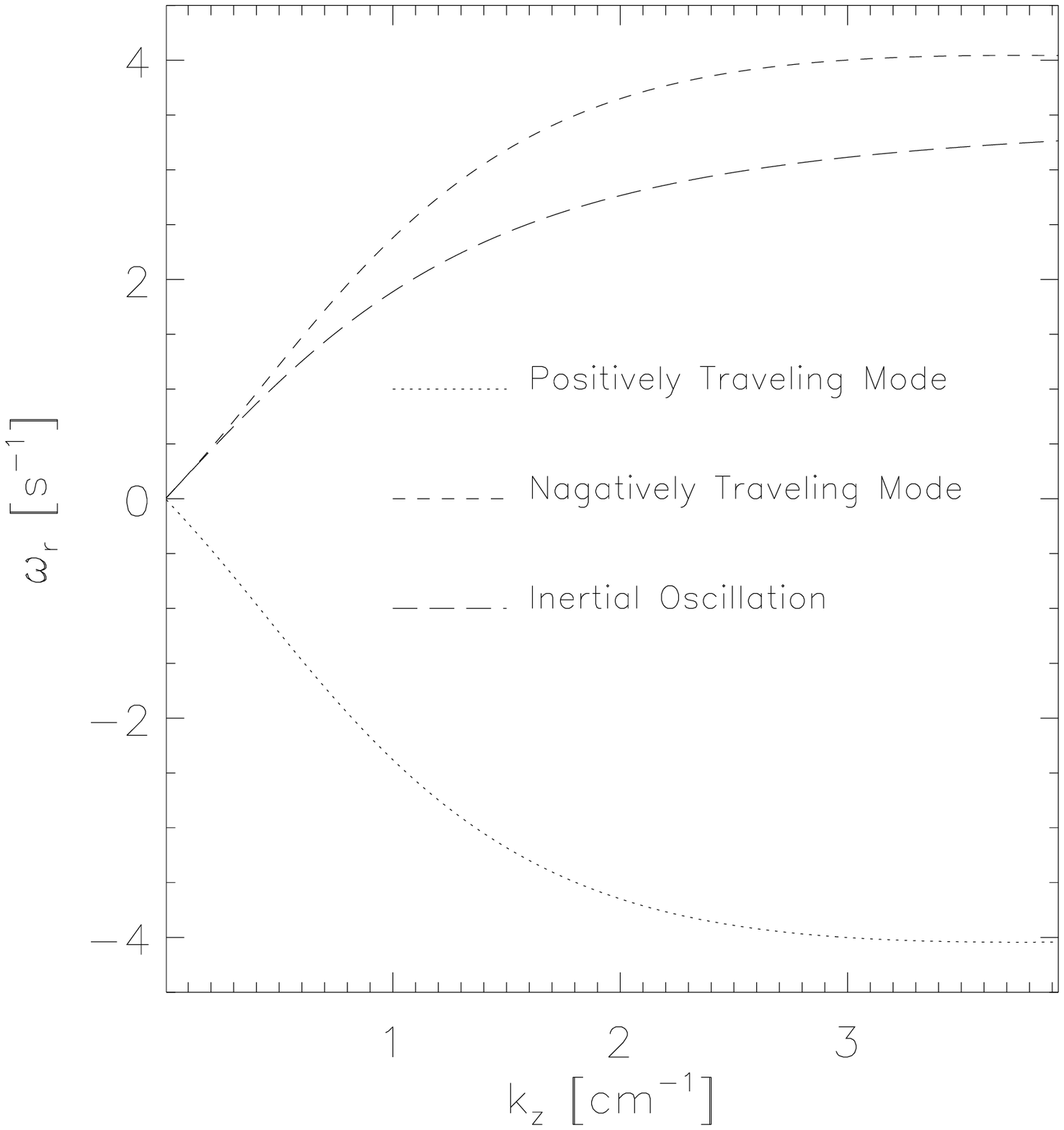}}}
\end{figure}

We conclude that \emph{in highly resistive flow, HMRI reduces to a
weakly destabilized inertial oscillation}.  In the present inviscid
approximation, instability persists to arbitrarily large resistivity,
though with reduced growth rate.  Furthermore, we note from
\eqref{roots} that instability [\emph{i.e.} $\Im(\omega)>0$] occurs
only if the bivalent signs are chosen so that $\Omega B_\theta B_z
k_z/\Re(\omega)<0$, which implies that \emph{the unstable mode
propagates axially with the same sense as the background Poynting
flux}.  [From \eqref{IO}, the group velocity $\pd\Re(\omega)/\pd k_z$
and phase velocity $\Re(\omega)/k_z$ have the same sign.]  Although we
have derived this propagation rule in the resistive limit, numerical evidence indicates
that it is true of the full dispersion relation \eqref{fulldisp},
as demonstrated by Figure.~\ref{dispersion}.

Instability requires the square brackets in \eqref{disp2} to be positive, whence
\[
2(\mu\wt)^2~\pm\frac{2+Ro}{\sqrt{1+Ro}}\wz(\mu\wt)~+\wz^2~<~0.
\]
The inequality is possible if and only if
the discriminant of the lefthand side, regarded as a quadratic
equation in $\mu\wt$, is positive:
\[
\frac{(2+Ro)^2}{1+Ro}\wz^2 ~-8\wz^2>0,
\]
which translates to
\begin{equation}\label{Rolimits}
Ro<2(1-\sqrt{2})\approx -0.8284\quad\mbox{\underline{or}}
\quad Ro>2(1+\sqrt{2})\approx 4.8284
\end{equation}
Thus, at least in WKB, \emph{the keplerian value $Ro=-3/4$ is
excluded, as of course is uniform rotation ($Ro=0$)}.  We say ``of course''
because, the background being current free, the only source of free energy
is the shear.


%
%
%

\subsection{Numerical results for wide gaps in periodic cylinders}
\label{sec:periodic}

We have adapted a code developed by \cite{gj02} to allow for a helical
field.  Vertical periodicity is assumed, but the radial equations are
solved directly by finite differences with perfectly conducting
boundary conditions.  The underlined terms in 
eqs.~(\ref{polind})-(\ref{toreuler})
are retained, and viscous terms are added although their influence is small
at Reynolds numbers of interest.
The code reproduces
published results for marginal stability \cite{rhss05,hr05}.
Table~\ref{table1} compares the predictions of the WKB dispersion
relation \eqref{fulldisp} with those of this code (labeled
``Global'').  The agreement is reasonably good, considering the
crudeness of the WKB approximation.  No unstable modes are found for
the parameters of Figure~{\ref{dispersion} at $Ro(r_{1})\ge -0.821$:
the keplerian value $Ro=-0.75$ is stable.

Astrophysical disks correspond to very wide gaps, $r_2-r_1\gg h$, as
well as keplerian rotation.  Given $(\Rm, S)=(0.1,0.03)$ and
$r_{2}/r_{1}=2.0,\;2.83,\;5.0$, the maximum unstable Rossby numbers at
the inner cylinder are found to be $Ro(r_{1})=-0.928$, $-0.947$, and
$-0.981$, respectively, from our radially global linear code.  We
conjecture that keplerian flows---more precisely, flows in which
$0\ge Ro\ge-3/4$ at all radii---are stable for all gap widths.  It
would be interesting to prove this.

We have also estimated a few growth rates with our nonlinear,
compressible non-ideal MHD code \cite{lgj06}, which is a modified
version of the astrophysical code ZEUS2D \cite{sn921}.  In this case,
we use the wide-gap geometry of the Princeton MRI experiment
\cite{jgk01,gj02}, except that the computation uses periodic vertical
boundaries: $r_{1}=7.1\cm$, $r_{2}=20.3\cm$, $h=27.9\cm$,
$\Omega_{1}=400\rpm$, $\Omega_{2}=53.3\rpm$, $B_{z}=500\G$,
$B_\theta(r_1)=1\kG$; the material properties are again based on
gallium: $\eta\approx2000\cm^2\s^{-1}$, $\nu\approx
3\times10^{-3}\cm^2\s^{-1}$. The growth rate and real frequency from
the ZEUS2D simulations are respectively $1.06 \unit{s^{-1}}$ and $3.93
\unit{s^{-1}}$, compared to $1.05 \unit{s^{-1}}$ and $3.89
\unit{s^{-1}}$ from the linear code.  WKB yields $(\gamma,\omega_r)=
(0.41,3.90)\unit{s^{-1}}$---not an accurate result for the growth
rate, but considering the width of the gap, the agreement is pleasing.

The growth rates in Table~\ref{table1} are of order $1\s^{-1}$, as
compared to $\sim30\s^{-1}$ for SMRI in this geometry at
the full rotation rate and field planned for 
the Princeton experiment \cite{lgj06}:
$\Omega_{1}=4,000\rpm$, $\Omega_{2}=533\rpm$, $B_{z}=5\kG$, and
$B_\theta=0$.  We have found this to be typical: the growth rates
of HMRI are small, at least in regimes where SMRI is stable.

 \begin{table}[!htp]

\caption{\label{table1} Comparison between WKB and numerical growth rates
in a vertically periodic Couette flow with the parameters of 
Figure~\ref{dispersion} except for a nonzero viscosity like that of gallium:
$\nu=3.1\times 10^{-3}$. The mode number $n\equiv k_z h/\pi$.}

\begin{tabular}{|c|c|c|c|c|}
\hline 
n&
WKB $\gamma$ [$s^{-1}$]&
WKB $\omega_{r}$ [$s^{-1}$]&
Global $\gamma$ [$s^{-1}$]&
Global $\omega_{r}$ [$s^{-1}$]\tabularnewline
\hline
\hline 
1&
0.1612&
0.9443&
0.0965&
1.4004\tabularnewline
\hline 
2&
0.3911&
1.9182&
0.3465&
2.5164\tabularnewline
\hline 
3&
0.5878&
2.7084&
0.6031&
3.2638\tabularnewline
\hline 
4&
0.7387&
3.2646&
0.7907&
3.7094\tabularnewline
\hline
5&
0.8356&
3.6221&
0.8960&
3.9549\tabularnewline
\hline
6&
0.8805&
3.8366&
0.9339&
4.0799\tabularnewline
\hline
7&
0.8829&
3.9565&
0.9241&
4.1352\tabularnewline
\hline
8&
0.8543&
4.0166&
0.8831&
4.1512\tabularnewline
\hline
9&
0.8049&
4.0400&
0.8227&
4.1451\tabularnewline
\hline

\end{tabular}
\end{table}

\subsection{Finite cylinders: a perturbative approach}\label{sec:insu}

In finite nonperiodic cylinders with insulating or partially
insulating endcaps, the MHD eigenfunctions are intrinsically two
dimensional: they are not separable in $r$ and $z$.  (Separability
could be achieved with perfectly conducting endcaps, but then the
axial field would be attached to them.  This would allow the boundary
to exert magnetic forces on the fluid, which seems undesirable and in
any case is experimentally less realistic than insulating endcaps.)
The purely hydrodynamic problem for $\eta=\infty$ \emph{is} separable,
however, if viscosity is neglected so that we may assume no-slip
boundary conditions.  This suggests a perturbative expansion of the
eigenvalue problem in $\eta^{-1}$---more properly,
$(\Rm,S)\to(\epsilon\Rm,\epsilon S)$, with $\epsilon$ a small
parameter.  The cylinders themselves are assumed infinitely long and
perfectly conducting; although this is not realistic, it does not
result in any attachment of the field to the boundaries, and it allows
the magnetic field more easily to be matched onto vacuum solutions
that decay as $|z|\to\infty$ in the regions above and below the fluid.
The underlined terms in equations \eqref{polind}-\eqref{toreuler} will
be neglected because they contribute to the eigenfrequency only at
$O(\eta^{-2})$ and higher orders.

We begin with the zeroth-order problem, $\emph{i.e.}$ for $\eta=\infty$.
As noted above, the hydrodynamic boundary conditions
\begin{equation}
\label{inobc}
\phi=0~\mbox{on}~r=r_1,r_2~\mbox{and on}~z=0,h,
\end{equation}
and inertial-mode equation \eqref{ino0} are separable, so we look for an
eigenmode of the form
\begin{equation}
  \label{phisep}
  \phi(t,r,z)= e^{-i\omega t}\varphi(r)\sin kz,\qquad k=n\frac{\pi}{h}\,.
\end{equation}
The radial function $\varphi(r)$ satisfies
\begin{equation}
  \label{radialeqn}
\frac{d^2\varphi}{dr^2}+\frac{1}{r}\frac{d\varphi}{dr}+\left[k^2
\left(\frac{4\ab^2}{\omega^2}-1\right)+\frac{1}{r^2}
\left(\frac{4\ab\bp k^2}{\omega^2}-1\right)\right]\varphi=0,
\end{equation}
assuming a Couette profile $\Omega(r)=\ab+\bp r^{-2}$ so that
$\kappa^2=4\ab\Omega$, which is satisfied by the Bessel functions
$J_\nu(pr)$ \& $Y_\nu(pr)$ if
\begin{equation}
  \label{bessels}
  \nu^2\equiv 1-\frac{4\ab\bp k^2}{\omega^2}\,,\qquad
p^2\equiv k^2\left(\frac{4\ab^2}{\omega^2}-1\right).
\end{equation}
We may thus solve this problem exactly. However, for qualitative
information, we notice that if we multiply (\ref{radialeqn}) by $r$ it
becomes 
\[
\frac{d}{dr}\left( r\frac{d\varphi }{dr}\right) +\left[ \frac{1}{\omega ^{2}}%
\left( 4\mathrm{a}^{2}k^{2}r+\frac{4\mathrm{a}\mathrm{b}k^{2}}{r}\right)
-\left( \frac{1}{r}+k^{2}r\right) \right] \varphi =0.
\]%
This is the same form as the Sturm-Liouville problem 
\[
\frac{d}{dr}\left( P(r)\frac{d\varphi }{dr}\right) +\left[ \lambda
R(r)-Q\left( r\right) \right] \varphi =0
\]%
\[
\varphi (r_{1})=0,\ \ \ \ \varphi (r_{2})=0,
\]%
where 
\begin{eqnarray*}
P(r) &=&r, \\
R(r) &=&4\mathrm{a}^{2}k^{2}r+\frac{4\mathrm{a}\mathrm{b}k^{2}}{r}>0, \\
Q(r) &=&k^{2}r+\frac{1}{r}>0, \\
\lambda  &=&1/\omega ^{2}.
\end{eqnarray*}%
Therefore $\lambda $ is real and positive (\cite%
[Chapter X]{Ince}); consequently the frequencies $\omega$ which we
seek are all real. There are no modes which grow in time. Thus we
conclude that, {\it all inviscid axisymmetric modes are neutrally
stable in the limit of infinite resistivity}. The coefficient
$R(r)=\Phi (r)$, the Rayleigh discriminant, so this result is to be
expected.

We may arrange for $\phi(r_1)=0$ by taking
\begin{equation}\label{phimn}
\varphi_{mn}(r)\equiv J_\nu(pr_1)Y_\nu(pr)-Y_\nu(pr_1)J_\nu(pr).
\end{equation}
Since we also require $\phi(r_2)=0$, the determinant
\begin{equation}
  \label{determ}
  \Delta(\omega,k)\equiv J_\nu(pr_1)Y_\nu(pr_2)-J_\nu(pr_2)Y_\nu(pr_1)
\end{equation}
must vanish.  The condition $\Delta=0$ defines a discrete set of
eigenfrequencies
$\omega_{1,n}>\omega_{2,n}>\ldots>\omega_{mn}\ldots>0$ for each
$k=k_n$.  Let $\phi_{m,n}$ be the complete eigenfunction
\eqref{phisep} corresponding to a given $k_n$ \& $\omega_{m,n}$. We define an inner product 
[here $\phi_{mn}$ is defined by \eqref{phisep} with $\varphi(r)\to\varphi_{mn}(r)$]
\begin{equation}
  \label{norm}
  \left\langle\phi_{m'n'},\phi_{mn}\right\rangle \equiv
\int\limits_0^h dz\int\limits_{r_1}^{r_2} rdr\,
\bar\phi_{m'n'}\phi_{mn},
\end{equation}
where the overbar denotes complex conjugation.  The eigenfunctions are orthogonal in the
sense that $\langle\phi_{mn},\kappa^2\phi_{m'n'}\rangle=0$
if $\omega_{mn}^2\ne\omega_{m'n'}^2$.

To get the $O(\eta^{-1})$ corrections to $\omega_{mn}$, we must
express the magnetic perturbations $\psi$ and
$b_\theta$ appearing on the righthand of \eqref{phi1} in terms of the
zeroth-order eigenfunctions $\phi_{mn}$.
Neglecting the time derivative in \eqref{polind} yields
\begin{equation}\label{bpsol}
D\psi=-\eta^{-1}B_z\pd_z\phi_{mn}.
\end{equation}
To get $b_\theta$ from
\eqref{torind}, we first use \eqref{toreuler} to write 
$v_\theta\approx (2\ab/i\omega_{mn})\pd_z\phi_{mn}$, so that
\begin{equation}
  \label{btsol}
  b_\theta\approx -2\eta^{-1}D_T^{-1}\left(\frac{B_\theta}{r}\pd_z\phi_{mn}
+\frac{iaB_zk_n^2}{\omega_{mn}}\phi_{mn}\right).
\end{equation}
Note that we have replaced $\pd_z^2$ with $-k_n^2$; we may similarly replace any
even power of $\pd_z$ but not an odd power, which changes a $\sin k_n z$ to
a multiple of $\cos k_n z$.
The operator $D_T^{-1}$ is the inverse of $D$
with the boundary conditions appropriate to $b_\theta$, which are different
from those of $\phi$ [eq.~\eqref{inobc}]:
\begin{equation}
  \label{btbcs}
  \pd_r^\dag b_\theta=0~\mbox{at}~r=r_1,r_2\quad
\mbox{and}~b_\theta=0~\mbox{at}~z=0,h.
\end{equation}
Using \eqref{bpsol} \& \eqref{btsol} to eliminate $D\psi$ and $b_\theta$
from \eqref{phi1} results in
\begin{multline}\label{perteqn}
\left(\pd_t^2 D+\kappa^2\pd_z^2\right)\phi=\\
-i\omega_{mn}\eta^{-1}\left[(k_nB_z)^2+4
\left(\frac{-iB_\theta}{r}\pd_z+\frac{\Omega B_z k_n^2}{\omega_{mn}}
\right)(-D_T^{-1})
\left(\frac{-iB_\theta}{r}\pd_z+\frac{\ab B_z k_n^2}{\omega_{mn}}\right)
\right]\phi_{mn}\,.
\end{multline}

On the righthand side of \eqref{perteqn}, the eigenmode and
eigenfrequency have been evaluated to zeroth order in $\eta^{-1}$.  On
the lefthand side, we must consider that
$\omega\to\omega_{mn}+\delta\omega$ and
$\phi\to\phi_{mn}+\delta\phi$, where $\delta\omega$ and $\delta\phi$
are of first order in $\eta^{-1}$.  
We may obtain an expression for $\delta\omega$ by taking the
inner product of \eqref{perteqn} with $\phi_{mn}$ and replacing
$i\pd_t\to\omega_{mn}+\delta\omega$ on the lefthand side.
The single term involving $\delta\phi$ at $O(\eta^{-1})$ is
$\langle\phi_{mn},(\kappa^2-\omega_{mn}^2D)\delta\phi\rangle$, and this
vanishes upon integration by parts.
On the right side, it is convenient
to define the self-adjoint operator
\begin{equation}\label{Hdef}
H\equiv 2\left(
-\frac{B_\theta}{r}i\pd_z+\frac{\ab B_z k_n^2}{\omega_{mn}}\right)
=H^\dag.
\end{equation}
At last, then,
\begin{multline}\label{firstorder}
-\left\langle\phi_{mn},\,D\phi_{mn}\right\rangle\delta\omega=\\
-\frac{i}{2\eta}\left[(k_n B_z)^2
\left\langle\phi_{mn},\phi_{mn}\right\rangle
-\left\langle H\phi_{mn},D_T^{-1}H\phi_{mn}\right\rangle
-\frac{2\bp B_z k_n^2}{\omega_{mn}}
\left\langle\phi_{mn},\,r^{-2}D_T^{-1}H\phi_{mn}\right\rangle
\right].
\end{multline}
Now $D$ and $D_T^{-1}$ are negative-definite operators.
Therefore, the only term that can make a positive contribution
to the growth rate $\Im(\delta\omega)$
is the last term on the righthand side, and specifically the part of $H$
involving $B_\theta \partial_z$ since $\ab\bp>0$.

To evaluate $\delta\omega$ from \eqref{firstorder}, we need
explicit expressions for $D$ and $D_T^{-1}$.  The first is
easy enough: it follows from \eqref{ino0} that 
$D\phi_{mn}=-(k_n^2\kappa^2(r)/\omega_{mn}^2)\phi_{mn}$.
For $D_T^{-1}$, we construct the eigenfunctions of $D$ with
the boundary conditions \eqref{btbcs}:
\begin{align}\label{chijn}
D\chi_{jn}(r,z)&= -(q_j^2+k_n^2)\chi_{jn}(r,z),\\
\label{chiform}
\chi_{jn}(r,z)&\equiv R_{jn}(r)\sin k_n z\,,\qquad k_n=n\frac{\pi}{h}
\end{align}
\begin{equation}
\mbox{where}~~ R_{jn}=
  \begin{cases}
  J_0(q_j r_1)Y_1(q_j r)-Y_0(q_j r_1)J_1(q_j r), 
  &\text{if $q_j \neq 0$;}\\
   r^{-1}
  &\text{if $q_{0} = 0$;}
  \end{cases}
\end{equation}
\begin{equation}
\label{qdet}
\mbox{and $q_j$ satisfies}~~
J_0(q_j r_1)Y_0(q_j r_2)-Y_0(q_j r_1)J_0(q_j r_2)\equiv 0
\end{equation}
When applied to $\chi_{jn}$,
$D_T^{-1}\to(q_j^2+k_n^2)^{-1}$.  An arbitrary function $f(r,z)$ can
be expanded in these eigenfunctions, so that
\begin{equation}\label{DTsol}
D_T^{-1}f(r,z)=-\sum_n\sum_j(q_j^2+k_n^2)^{-1}
\frac{\langle\chi_{jn},f\rangle}{\langle\chi_{jn},\chi_{jn}\rangle}\,
\chi_{jn}(r,z)\,.
\end{equation}
The important point is that $D_T^{-1}$ turns a function proportional
to $\sin k_n z$ into another such.  Therefore,
$\langle\phi_{mn},D_T^{-1}\pd_z\phi_{mn}\rangle=0$, and so the part
of $H$ involving $r^{-1}B_\theta i\pd_z$ does not contribute to
the expression \eqref{firstorder} for the first-order eigenfrequency.
This, however, was the only term that might have made for a positive
growth rate.  We conclude that {\it at $O(\eta^{-1})$, HMRI does not
grow in finite cylinders with insulating endcaps.}

The same perturbative method could have been used for
periodic vertical boundary conditions;
$\phi_{mn}$ and $\chi_{jn}$ would have involved $\exp(ik_z z)$ instead
of $\sin{k_n z}$.  The term involving
$r^{-1}B_\theta i\pd_z$ in eq.~\eqref{firstorder}
would then have contributed to the growth rate with the same sign as
$-(k_z/\omega_{mn})\Omega B_\theta B_z$.
Evaluating this term, we conclude that in highly resistive periodic flows, 
(i) unstable modes propagate axially in the direction of the
background Poynting flux---as found in WKB; and (ii)
the instability occurs only if $\beta> a k_z r/\omega_{mn}$ because
of the second term in $H$.

We have written MATLAB procedures to evaluate eq.~(\ref{firstorder}).
The results confirm our conclusions above.  When periodic boundary
conditions are used, the perturbative result matches the growth
rate found from our radially global linear code to three digits
in sufficiently resistive cases: \emph{e.g.,}
$\gamma=1.89\times10^{-3}\Omega_1$ in the Princeton
geometry with $\Rm=0.1$, $S=0.043$, $\Omega_2/\Omega_1=0.1325$, $\beta=2$.
But when insulating endcaps are imposed, the perturbative estimate
of the growth rate is always negative.

\subsection{Finite cylinders: two other approaches}\label{sec:guide}

Here we analyze finite cylinders by approximations that do not require
large resistivity: by a variant of WKB, and by direct axisymmetric numerical
simulations.

In the modified WKB approach, perturbations are again assumed to vary
as $\exp(ik_r+st)$ with a common complex growth rate $s\equiv-i\omega$
and radial wavenumber $k_r=\pi/(r_2-r_1)$, but the vertical dependence is treated
differently.  With the $t$ and $r$ dependence factored out, the
linearized equations of motion reduce to homogeneous ordinary
differential equations with coefficients independent of $z$.
Elementary solutions of these equations exist with exponential dependence on
$z$; however, since the vertical boundaries are not
translationally invariant, the wavenumber $k_z$ need not be real, and
growing modes can be linear combinations of the elementary
exponential solutions with the same $\omega$ but different $k_z$.
The vertical magnetic boundary conditions require the fields to match
onto a vacuum solutions that decay exponentially as $|z|\to\infty$ in
the space $r_1\le r\le r_2$ between the extended conducting cylinders:
\begin{equation}
  \label{wgbcs}
  z=0:\quad \phi=b_\theta=0,~\pd_z\psi=|k_r|\psi;\qquad
  z=h:\quad \phi=b_\theta=0,~\pd_z\psi=-|k_r|\psi.
\end{equation}

We search iteratively for such modes as follows.  Given a trial value
for $s$, the dispersion
relation \eqref{fulldisp} has six roots---in general complex---for
the vertical wavenumber, which can be regarded
as algebraic functions of the growth rate: $\{k_{z,\alpha}(s)\}$,
 $\alpha\in\{1,\ldots,6\}$.
We seek a mode in the finite cylinder of the form
\begin{equation}
  \label{eq:combo}
  \b{q}(t,r,z)\equiv [\phi,v_\theta,\psi,b_\theta]^{T}=
~e^{st+ik_r r}\sum\limits_{\alpha=1}^6 Y_\alpha
\b{q}_{\alpha}
\exp(ik_{z,\alpha}z).
\end{equation}
Each term in the sum above is the elementary solution corresponding
to a particular root $k_{z,\alpha}(s)$, with $\b{q}_\alpha$ a 4-component
column vector; these elementary solutions are superposed with constant weights
$\{Y_\alpha\}$.
Substitution into the boundary conditions \eqref{wgbcs} yields
a sixth-order homogeneous linear system for the $\{Y_\alpha\}$.
Nontrivial solutions exist only if the determinant $\mathcal{D}(s)$ of
this system vanishes.  The equation $\mathcal{D}(s)=0$ is transcendental
and we cannot solve it analytically, but a numerical
nonlinear zero-finding algorithm recovers the roots for $s$.

We have checked this procedure by replacing \eqref{wgbcs} with periodic
boundary conditions and comparing the results with direct solutions of
the dispersion relation \eqref{fulldisp}.  Also, we find reasonably
good agreement with growth rates determined from ZEUS2D
simulations of a narrow-gap configuration with insulating
boundaries (see below).
However, for sufficiently large resistivity, no roots with positive
$\Re(s)$ are found, in agreement with the perturbative results of
Section \ref{sec:insu}.

For the ZEUS2D simulations, we represent the poloidal magnetic field
at $z\le0$ and $z\ge h$ by flux functions $\Phi_\pm(r,z)$ satisfying
$b_r\b{e}_r+b_z\b{e}_z=r^{-1}\b{e}_\theta\times\b{\nabla}\Phi$ and
$\b{\nabla\times b}=0$.  The latter implies
$r\pd_r(r^{-1}\pd_r\Phi)+\pd_z^2\Phi=0$,
which is solvable by separation
of variables since we require $\Phi=0$ on the vertically extended
conducting cylinders.  The elementary solutions are
\[
  \Phi_k(r,z)\propto
re^{-k|z-z_0|}\left[Y_1(kr_1)J_1(kr)-J_1(kr_1)Y_1(kr)\right]\,,
\]
for an infinite discrete set of nonnegative values of $k$ determined by
$\Phi_k(r_2,z)=0$.
At each endcap, we match the vertical field $b_z$ protruding from
the fluid with a superposition of vacuum solutions of this form, and
thereby obtain a boundary condition relating $b_z$ and $b_r$.  Of course
$b_\theta=0$ at these boundaries since the current along the axis is constant.

We have performed simulations with insulating endcaps
for the parameters of Figure~\ref{dispersion}.
We find a complex growth rate
$s\approx 0.51+4.18 i~\s^{-1}$, as compared to $s\approx 0.37+3.68
i~\s^{-1}$ from the modified WKB approach
\eqref{wgbcs}-\eqref{eq:combo} above.  Considering the approximate
nature of the latter approach, the agreement is satisfactory.
We have also carried out ZEUS2D simulations with insulating endcaps in
the wide-gap experimental geometry [$(r_1,r_2,h)=(7.1,20.3,28)\cm$].  Here we
find a growth rate $\sim 0.27\s^{-1}$, as opposed to $\sim 1.06\s^{-1}$ with
periodic boundaries.  We conclude that insulating endcaps lower the
growth rate, even in flows of moderate $(\Rm,S)$.

A limitation of our direct simulations is that since we use explicit
time stepping, we cannot explore very large resistivities
\citep{lgj06}.  The modified WKB approach does not suffer from any
restriction on $\eta$, but it is not trustworthy for wide gaps.  The
concordance between the two approaches where both are
applicable---namely for narrow gaps and moderate $(\Rm,S)$---inclines
us to trust results obtained from one of these approaches in regimes
where the other is not applicable.  In particular, the modified WKB
method predicts that highly resistive flows are completely stable in
finite cylinders, at least for narrow gaps.  The perturbative analysis
of Section \ref{sec:insu} reaches the same conclusion for gaps of any
width, but that analysis is valid at $O(\eta^{-1})$ only.

\section{Conclusions}\label{sec:dis}
We have analyzed the linear development
of helical magnetorotational instability in a non-ideal magnetohydrodynamic
Taylor-Couette flow, paying particular attention to
the effects of the axial boundary conditions.  A number of complementary
approximations and numerical methods have been used.

For infinitely long or periodic cylinders, we confirm that there is an
axisymmetric MHD instability that persists to smaller magnetic
Reynolds number and Lundquist number in the presence of \emph{both}
axial and toroidal background magnetic field than the standard MRI
that exists for axial field alone. The new mode is an overstability
and propagates axially
in the direction of the background Poynting flux
$-r\Omega B_\theta B_z/\mu_0$.  In highly resistive flows, the new
mode is a weakly destabilized hydrodynamic inertial oscillation.
Growth depends also on the ratio of shear to rotation, \emph{i.e.}
Rossby number: for all aspect ratios $r_2/r_1$ that we have explored,
and certainly for narrow gaps, the keplerian Rossby number is stable.
Even for those profiles that
permit growth, the rate tends to be rather small, except in flows that
are sufficiently ideal to permit growth of standard MRI (axial field only).

We have also considered finite cylinders with insulating endcaps,
which are closer to experimental reality but which do not
permit traveling modes that propagate indefinitely along the axis.
Astrophysical disks also, of course, have limited vertical thickness.
These boundary conditions reduce the growth rate of the helical mode,
and stabilize highly resistive flows entirely.
The small growth rate may make it difficult to detect the instability
in the face of Ekman circulation and other experimental imperfections.
In addition, the new mode requires toroidal fields at least as large
as the axial field, and therefore large axial currents, which
introduces additional engineering challenges. 

For all of these reasons, the experimental advantage of helical MRI
over standard MRI is open to question, as is the relevance of
HMRI to astrophysical disks, although it may be relevant to 
stellar interiors and jets, where the magnetic geometry and the Rossby number may be
more favorable. Also, HMRI may have theoretical
significance that goes beyond its direct applications.  It is
not understood why 
linearly and axisymmetrically stable rotating flows are
often also nonlinearly and nonaxisymmetrically unstable, especially
since subcritical transition does occur at some
Rossby numbers \citep{ll05}.
The fact that even a very poorly coupled magnetic field can sometimes
linearly destabilize such flows hints that it might also affect
nonlinear transition.

\acknowledgments
The authors would like to thank James Stone for the advice on the ZEUS
code. This work was supported by the US Department of Energy, NASA
under grants ATP03-0084-0106 and APRA04-0000-0152, the
National Science Foundation under grant AST-0205903, and in part by the U. S. Department of
Energy under Grant No. DE-FG02-05ER25666 (to I.H.).



\end{document}